\documentclass[aps,reprint,showpacs,twocolumn,superscriptaddress,floatfix]{revtex4-1}
\usepackage{epsfig}
\usepackage[T1]{fontenc}
\usepackage[utf8]{inputenc}
\usepackage{amsmath}
\usepackage{amssymb}
\usepackage{amsfonts}
\usepackage{mathptmx}
\usepackage{textcomp}
\usepackage{dcolumn}
\usepackage{eucal}
\usepackage{bm}
\usepackage{color}
\usepackage[colorlinks,linkcolor=blue,citecolor=blue]{hyperref}

\usepackage{epstopdf}

\begin{document}

%\bibliographystyle{prsty}%{apsrev}

%%%%%%%%%%%%%%%%%%%%%%%%%%%%%%%%%%%%%%%%%%%%%%%%%%%%%%%%%%%%
%%%%%%%%%%%%%%%%%%%%%%%%%%%%%%%%%%%%%%%%%%%%%%%%%%%%%%%%%%%%
\title{Nanoscale phase-engineering of thermal transport with a Josephson heat modulator}

\author{Antonio Fornieri}
\email{antonio.fornieri@sns.it}
\affiliation{NEST, Istituto Nanoscienze-CNR and Scuola Normale Superiore, Piazza S. Silvestro 12, I-56127 Pisa, Italy}

\author{Christophe Blanc}
%\email{christophe.blanc@nano.cnr.it}
\affiliation{NEST, Istituto Nanoscienze-CNR and Scuola Normale Superiore, Piazza S. Silvestro 12, I-56127 Pisa, Italy}

\author{Riccardo Bosisio}
\affiliation{SPIN-CNR, Via Dodecaneso 33, I-16146 Genova, Italy}
\affiliation{NEST, Istituto Nanoscienze-CNR and Scuola Normale Superiore, Piazza S. Silvestro 12, I-56127 Pisa, Italy}

\author{Sophie D'Ambrosio}
\affiliation{NEST, Istituto Nanoscienze-CNR and Scuola Normale Superiore, Piazza S. Silvestro 12, I-56127 Pisa, Italy}

\author{Francesco Giazotto}
\email{francesco.giazotto@sns.it}
\affiliation{NEST, Istituto Nanoscienze-CNR and Scuola Normale Superiore, Piazza S. Silvestro 12, I-56127 Pisa, Italy}

%%%%%%%%%%%%%%%%%%%%%%%%%%%%%%%%%%%%%%%%%%%%%%%%%%%%%%%%

\date{\today}% It is always \today, today,

%%%%%%%%%%%%%%%%%%%%%%%%%%%%%%%%%%%%%%%%%%%%%%%%%%%%%%%%%%%%
%%%%%%%%%%%%%%%%%%%%%   ABSTRACT         %%%%%%%%%%%%%%%%%%%
%%%%%%%%%%%%%%%%%%%%%%%%%%%%%%%%%%%%%%%%%%%%%%%%%%%%%%%%%%%%

%\begin{abstract}
%\end{abstract}


\maketitle

%\tableofcontents
%%%%%%%%%%%%%%%%%%%%%%%%%%%%%%%%%%%%%%%%%%%%%%%%%%%%%%%%%%%%
%%%%%%%%%%%%%%%%%%%%   INTRODUCTION   %%%%%%%%%%%%%%%%%%%%%%
%%%%%%%%%%%%%%%%%%%%%%%%%%%%%%%%%%%%%%%%%%%%%%%%%%%%%%%%%%%%

\textbf{Macroscopic quantum phase coherence has one of its pivotal expressions in the Josephson effect~\cite{Josephson}, which manifests itself both in charge~\cite{Tinkham} and energy transport~\cite{MakiGriffin,GiazottoNature,MartinezNature}.
The ability to master the amount of heat transferred through two tunnel-coupled superconductors by tuning their phase difference is the core of coherent caloritronics~\cite{GiazottoNature,MartinezNature,MartinezRev}, and is expected to be a key tool in a number of nanoscience fields, including solid state cooling~\cite{GiazottoRev}, thermal isolation~\cite{MartinezNatRect,FornieriRev}, radiation detection~\cite{GiazottoRev}, quantum information~\cite{NielsenChuang,Spilla} and thermal logic~\cite{LiRev}. Here we show the realization of the first balanced Josephson heat modulator~\cite{MartinezAPLdouble} designed to offer full control at the nanoscale over the phase-coherent component of thermal currents. Our device provides magnetic-flux-dependent temperature modulations up to 40 mK in amplitude with a maximum of the flux-to-temperature transfer coefficient reaching 200 mK per flux quantum at a bath temperature of 25 mK. Foremost, it demonstrates the exact correspondence in the phase-engineering of charge and heat currents, breaking ground for advanced caloritronic nanodevices such as thermal splitters~\cite{Bosisio}, heat pumps~\cite{Ren} and time-dependent electronic engines~\cite{Valenzuela,Campisi,Niskanen,Quan}.}

When two superconductors (S$_1$ and S$_2$) are coupled by means of a thin insulating layer (I), they form a Josephson junction (JJ) through which a dissipationless \textit{charge} current $I_{\rm J}$ can flow due to Cooper pair tunneling. $I_{\rm J}$ obeys the well-known expression~\cite{Josephson}
\begin{equation}
I_{\rm J}=I_0\;\rm sin \varphi,
\end{equation}
where $I_0$ is the critical current of the JJ and $\varphi$ the phase difference between S$_1$ and S$_2$. Although the Cooper pair condensate carries no entropy under static conditions and the supercurrent $I_{\rm J}$ cannot contribute to the heat current~\cite{MakiGriffin,Guttman}, it was shown~\cite{MakiGriffin,GiazottoNature,MartinezNature} that the Josephson effect has a thermal counterpart. Indeed, if we impose a temperature bias to the JJ by setting the temperature in S$_1$ to be $T_1>T_2$ ($T_2$ being the temperature in S$_2$), a finite stationary \textit{thermal} current $J_{\rm tot}$ will flow through the junction~\cite{MakiGriffin,Guttman,GiazottoAPL},
\begin{equation}
J_{\rm tot}(T_1,T_2)=J_{\rm qp}(T_1,T_2)-J_{\rm int}(T_1,T_2)\rm\; cos \varphi.\label{Jtot}
\end{equation} 
\noindent In Eq.~\eqref{Jtot}, the term $J_{\rm qp}$ accounts for the energy transferred by quasiparticles whereas $J_{\rm int}$ represents the phase-dependent component of the heat current which originates from energy-carrying tunneling processes involving transfer of Cooper pairs~\cite{MakiGriffin,Guttman} (see Methods for further details).
\begin{figure}
\centering
%\hspace*{-2.em}
\includegraphics[width=\columnwidth]{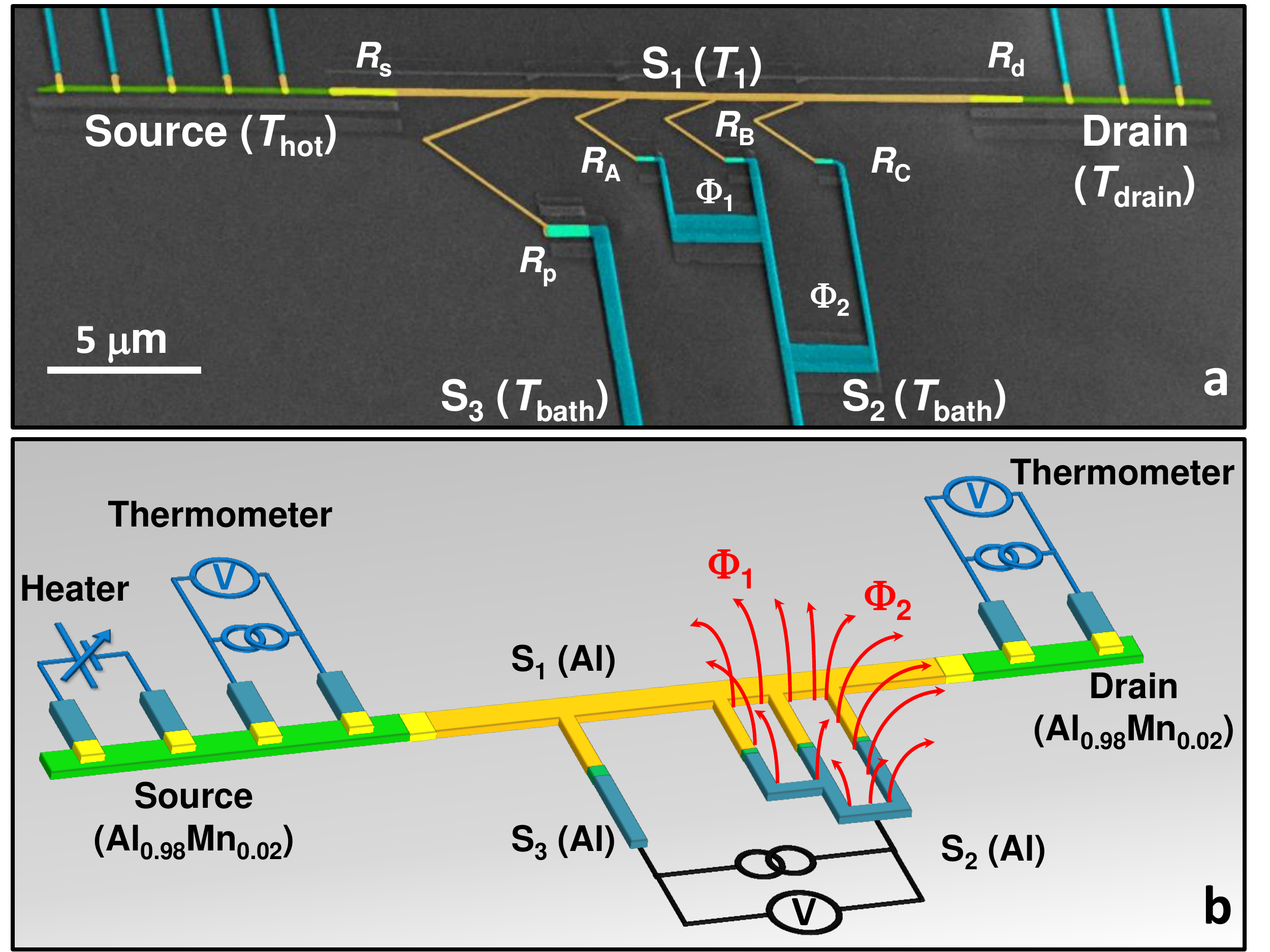}
%\vspace*{-4.ex}
\caption{\textbf{Quantum modulator structure.} \textbf{a.} Pseudo-color scanning electron micrograph of the double-loop heat interferometer. Source and drain N electrodes, depicted in green, are composed of Al$_{0.98}$Mn$_{0.02}$ and are connected to the Al island S$_1$ (represented in orange) and to a set of Al wires (dark cyan) serving either as heaters or thermometers. S$_1$ is tunnel-coupled to S$_2$ (Al, dark cyan) by means of three parallel JJs forming the double-loop SQUID and to a superconducting probe S$_3$ (Al, dark cyan). All the junctions in the structure are implemented through AlO$_{\rm x}$ tunnel barriers. The areas of the loops are of about 15 \textmu m$^2$ and 30 \textmu m$^2$. 
%In red, the source and drain are layers of 15 nm thick aluminium manganese. In Yellow the superconducting island is made of 20 nm of Al. The lower parts of the SQUID and the probe (blue) are also Al (40 nm thick). The green parts are oxidized tunneling junction. Read the methods for further details of the fabrication. 
\textbf{b.} Schematic picture of the device. The setup for electrical and thermal measurements are represented in black and blue, respectively, while the red lines indicate the magnetic fluxes $\Phi_1$ and $\Phi_2$ piercing the loops.\label{Fig1}}
%% \vspace*{-3.ex}
\end{figure}

Here we experimentally demonstrate that a double-loop direct current superconducting quantum interference device (DC SQUID)~\cite{SQUIDhandbook} with three parallel JJs~\cite{MartinezAPLdouble} represents a fundamental building block to achieve full phase-engineering of electronic heat currents at the nanoscale. Our device is robust against structure asymmetries, providing an almost complete annihilation (up to 99\%) of $J_{\rm int}$, whereas a conventional single-loop DC SQUID would suffer from a reduced contrast in the modulation. Yet, it allows to manipulate at will the heat current-phase relation, enabling the generation of exotic thermal interference patterns with large modulation amplitude, and enhanced sensitivity to magnetic flux variations~\cite{MartinezAPLdouble}.

The double-loop Josephson heat modulator has been fabricated by electron-beam lithography, three-angle shadow-mask evaporation of metals and \textit{in situ} oxidation (see Methods). It basically consists of five different parts, as shown in Fig.~\ref{Fig1}a. Aluminum (with a critical temperature $T_{\rm c}\sim 1.4$ K) was used to implement all the superconducting elements of the structure, while Al$_{0.98}$Mn$_{0.02}$ was employed as the normal metal (N)~\cite{MartinezNature,MartinezNatRect,Maasilta}. The source and drain N electrodes act as thermal reservoirs and are tunnel-coupled to a superconducting island (S$_1$) through junctions with normal-state resistances $R_{\rm s}\simeq 5.6$ k$\Omega$ and $R_{\rm d}\simeq 6.25$ k$\Omega$, respectively. S$_1$ defines the upper branch of the heat modulator and is connected to the lower branch (S$_2$) by means of three tunnel junctions forming the SQUID characterized by an equivalent resistance $R_{\rm SQUID}\simeq 700 \; \Omega$. The ratio between the areas of the interferometer loops is $\sim 2$. A superconducting probe (S$_3$) is also connected to S$_1$ via a tunnel-junction with a normal-state resistance $R_{\rm p} \simeq 600 \; \Omega$. Finally, the source and drain electrodes are tunnel-coupled to a set of external superconducting wires, forming NIS junctions of area $\sim 150\times 200$ nm$^2$ and normal-state resistance of $\sim 20$ k$\Omega$ each, which can be used as Joule heaters and thermometers~\cite{GiazottoRev}. 

\begin{figure}
\centering
%\hspace*{-2.em}
\includegraphics[width=\columnwidth]{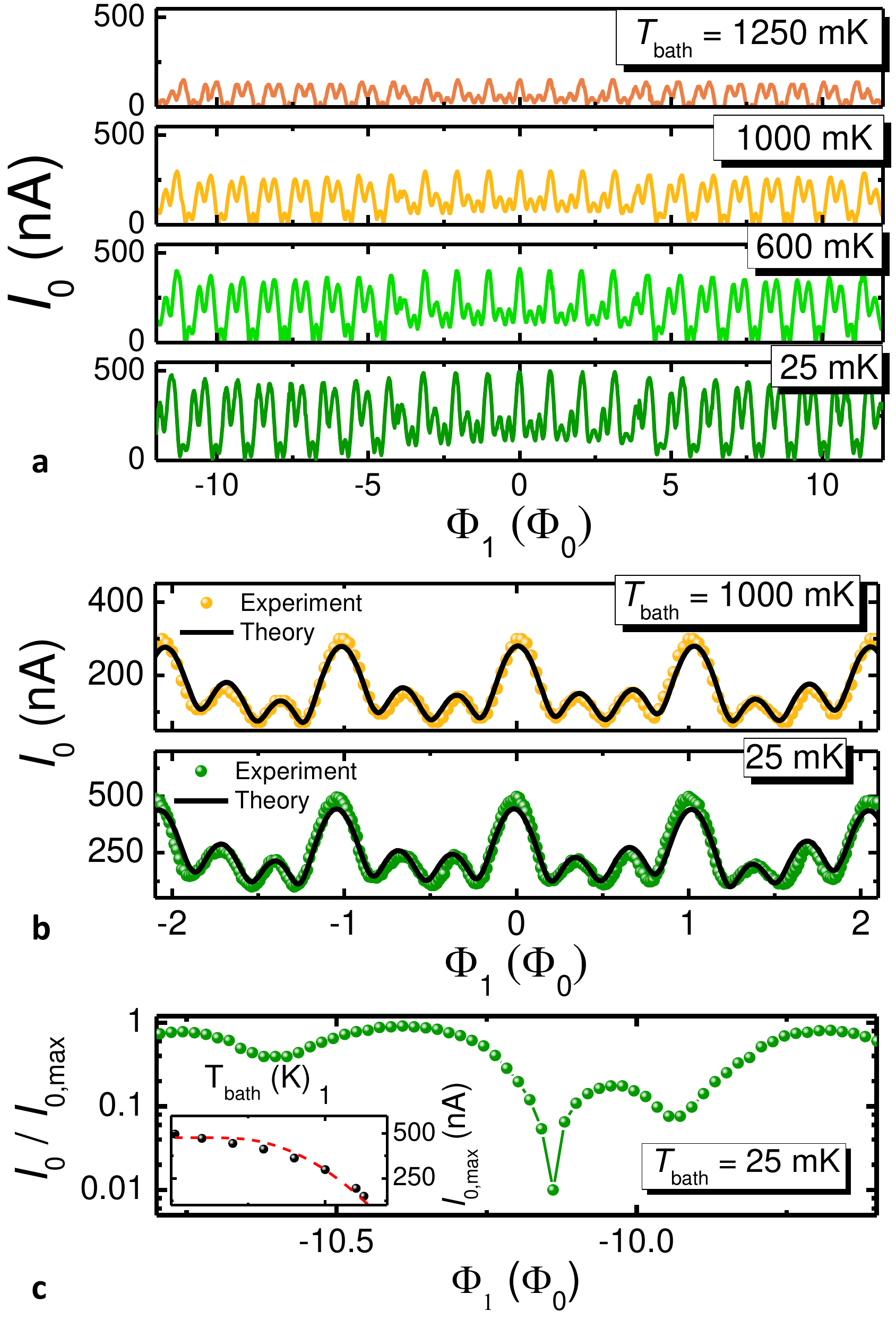}
%\vspace*{-4.ex}
\caption{\textbf{Electrical response of the interferometer.} \textbf{a.} SQUID total critical current $I_0$ vs magnetic flux $\Phi_1$ piercing the smallest of the interferometer loops for selected values of the bath temperature $T_{\rm bath}$. \textbf{b.} Zoom of two $I_0(\Phi_1)$ characteristics measured at $25$ mK and $1$ K. The full circles are experimental values, whereas the solid black lines are theoretical results obtained from Eq.~\eqref{ictheory}. \textbf{c.} Normalized critical current $I_0/I_{\rm 0,max}$ behaviour in a selected range of $\Phi_1$ for $T_{\rm bath}=25$ mK. The maximum observed suppression of $I_0$ is $\sim 99$ \%. The inset shows the maximum critical current $I_{\rm 0,max}=I_0(\Phi_1=0)$ (filled symbols) as a function of $T_{\rm bath}$ together with the theoretical Ambegaokar-Baratoff expectation (red dashed line, see text for details). \label{Fig2} }
%% \vspace*{-3.ex}
\end{figure}

To prove the perfect correspondence in the modulations of charge and heat currents, we performed two kinds of measurements, whose schematics are displayed in Fig.~\ref{Fig1}b. First, we investigated the dissipationless electrical transport across the interferometer via the superconducting electrode S$_3$ with a conventional four-wire technique. The junction between S$_1$ and S$_3$ was specifically designed to have a larger critical current than the SQUID one, thereby allowing an exact determination of the interferometer critical current. Voltage-current characteristics were recorded for different values of the magnetic-flux threading the loops. A well-defined Josephson current was observed with a maximum critical current $I_{\rm 0,max}\simeq 495$ nA at a bath temperature $T_{\rm bath}=25$ mK. The magnetic-flux interference pattern of the SQUID critical current $I_0$ for selected values of $T_{\rm bath}$ is shown in Fig.~\ref{Fig2}a. Its behavior can be understood by writing the expression of the total Josephson current flowing through the SQUID and imposing flux quantization constraints for both loops~\cite{Tinkham}:
\begin{equation}
%\begin{split}
\begin{aligned} 
I_{\rm J, SQUID} &= I_{\rm A} \sin(\varphi_{\rm A})+I_{\rm B} \sin(\varphi_{\rm B})+I_{\rm C} \sin(\varphi_{\rm C}), \\
\varphi_{\rm A} &= \varphi_{\rm B}  - \frac{2\, \pi  \, \Phi_{1}}{\Phi_{0}}+2n\pi, \\
\varphi_{\rm C} &= \varphi_{\rm B} + \frac{2\, \pi  \, \Phi_{2}}{\Phi_{0}}+2n\pi.
\label{Squidsupercurrent}
\end{aligned}
%\end{split}
\end{equation}
Here, $I_{\rm i}$ and $\varphi_{\rm i}$ are the critical current and the phase difference in the i-th JJ (being $\rm i=A,B,C$). $\Phi_{1}$ and $\Phi_{2}$ are the magnetic fluxes in the loops, while $\Phi_{0}\simeq 2 \times 10^{-15}$ Wb is the superconducting flux quantum.

For given $\Phi_1$ and $\Phi_2$, $I_0$ is obtained by maximizing $I_{\rm J,SQUID}$ with respect to $\varphi_{\rm B}$, leading to the following expression:
\begin{equation}
\label{ictheory} 
\begin{aligned}
I_{0}&=I_{\rm B} \sqrt{1+r_{1}^{2}+r_{2}^{2}+2 r_{1}\alpha + 2r_{2}\beta + 2 r_{1}r_{2}\gamma}\\
 &= I_{\rm B} \cdot \mathcal{F}_{int}(\Phi_1,\Phi_2), 
\end{aligned} 
\end{equation}
where $\alpha=\cos(2\pi\Phi _{1}/\Phi_{0})$, $\beta=\cos(2\pi \Phi _{2}/\Phi_{0})$, $\gamma=\cos[2\pi (\Phi _{1}+\Phi _{2})/\Phi_{0}]$, $r_{1}=I_{\rm A}/I_{\rm B}=R_{\rm B}/R_{\rm A}$ and  $r_{2}=I_{\rm C}/I_{\rm B}=R_{\rm B}/R_{\rm C}$ (see Fig.~\ref{Fig1}a).

\begin{figure*}%[t]
\centering
%\hspace*{-2.em}
\includegraphics[width=0.8\textwidth]{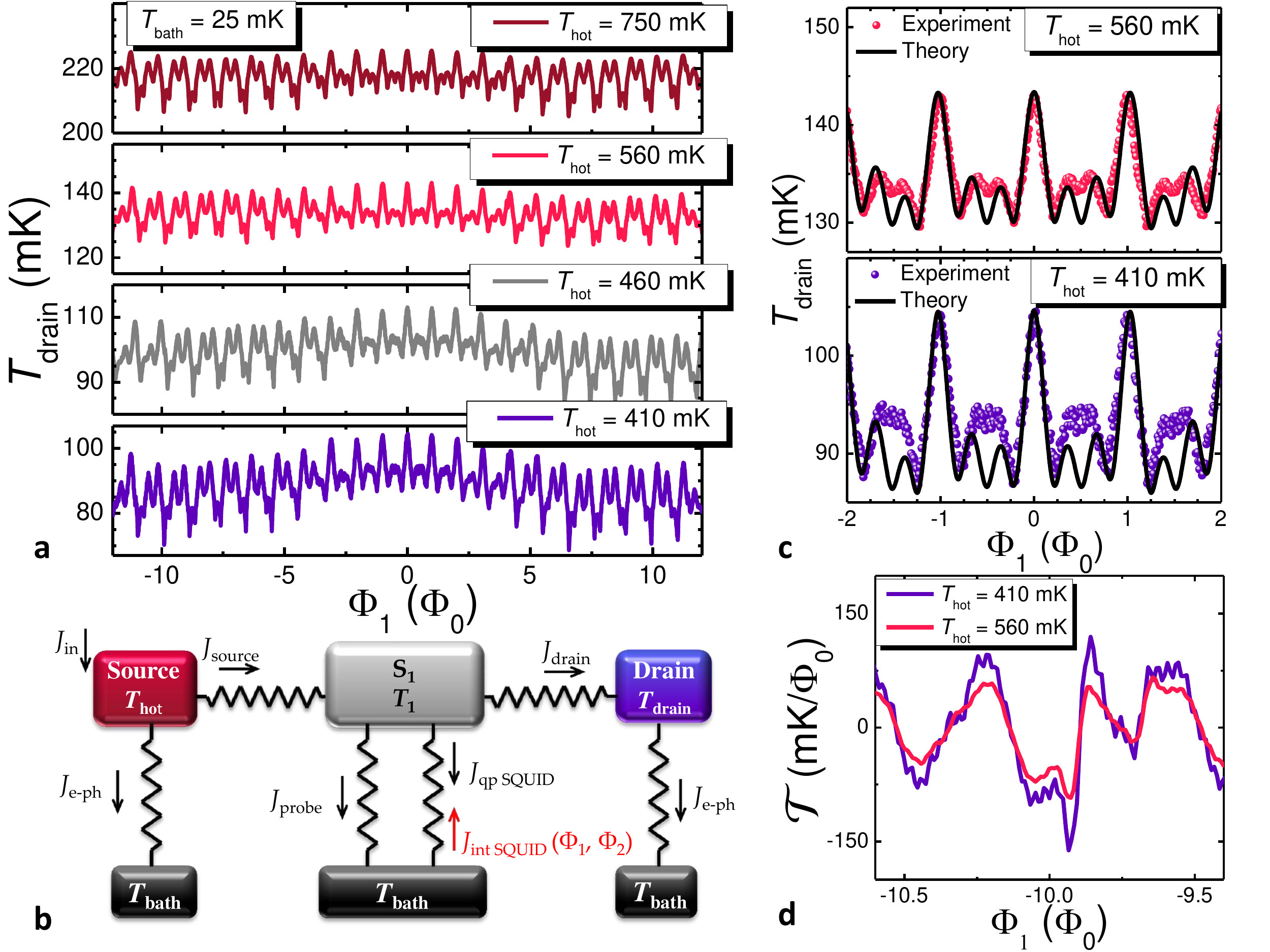}
%\vspace*{-4.ex}
\caption{\textbf{Thermal response of the Josephson heat modulator at 25 mK.} \textbf{a.} Magnetic flux modulation of the drain temperature $T_{\rm drain}$ for several values of the source temperature $T_{\rm hot}$. \textbf{b.} Thermal model outlining the relevant heat exchange mechanisms present in our interferometer. Arrows indicate heat current directions for the operating device, i.e., when $T_{\rm hot}> T_1> T_{\rm drain}> T_{\rm bath}$ (see text). The only existing magnetic-flux-dependent thermal current is $J_{\rm int,SQUID}$ (labelled in red). \textbf{c.} Experimental $T_{\rm drain}$ pattern (filled circles) around $\Phi_1=0$ for two selected values of $T_{\rm hot}$ along with the theoretical results from the thermal model (solid black lines). \textbf{d.} Flux-to-temperature transfer function $\mathcal{T}$ versus $\Phi_1$ for the same values of $T_{\rm hot}$ as in panel c.\label{Fig3}}
%% \vspace*{-3.ex}
\end{figure*} 
 
Figure~\ref{Fig2}b shows the excellent agreement between the experiment and the model, which allows us to extract precise values of the interferometer structural parameters. From the theoretical fits we obtain $\Phi_{2}=(1.93 \pm 0.02)\; \Phi_{1}$, $r_{1}=(0.93 \pm 0.04)$ and $r_{2}=(0.70 \pm 0.03)$. These values are consistent with the geometrical parameters expected from scanning electron micrograph of the device (see Fig.~\ref{Fig1}). The temperature-dependence of $I_{\rm 0,max}$ is plotted in the inset of Fig.~\ref{Fig2}c along with the Ambegaokar-Baratoff prediction~\cite{AB} for a JJ with a normal-state resistance $R_{\rm SQUID}$ and $T_{\rm c}\simeq 1.4$ K. The good agreement between theory and experiment confirms the ideal behaviour of our double-loop Josephson heat modulator. Moreover, as shown in Fig.~\ref{Fig2}c, an optimal suppression of the critical current of $\simeq 99$ \% is obtained for $|\Phi_1|\sim 10\; \Phi_0$, highlighting the excellent phase control of the Josephson current offered by our device.  

On the thermal side, the interferometric properties of our heat modulator were investigated by imposing a temperature gradient across the device. We emphasize that our experiment is focused on the heat carried by electrons only. We assume that lattice phonons present in every part of our structure are fully thermalized with the substrate phonons residing at $T_{\rm bath}$, thanks to the vanishingly small Kapitza resistance between thin metallic films and the substrate at low temperatures~\cite{GiazottoNature, MartinezNature,MartinezNatRect,Wellstood}. When we inject a Joule power ($J_{\rm in}$) through a couple of superconducting probes connected to the source electrode (see Fig.~\ref{Fig1}a), we can raise its electronic temperature $T_{\rm hot}$ significantly above $T_{\rm bath}$~\cite{Wellstood}. The quasiparticle temperature $T_1$ in S$_1$ is therefore increased, yielding a thermal gradient across the SQUID. This hypothesis is expected to hold because the lower branch of the interferometer (S$_2$) extends into a large-volume lead, providing efficient thermalization of its quasiparticles at $T_{\rm bath}$. This thermal gradient gives rise to the appearance of a finite heat current $J_{\rm tot}$ flowing through each JJ of the SQUID, leading to a periodic magnetic modulation of the drain electronic temperature ($T_{\rm drain}$). Both $T_{\rm hot}$ and $T_{\rm drain}$ are monitored by measuring the voltage drop across two pairs of current-biased NIS junctions~\cite{GiazottoRev}, as shown in Fig.~\ref{Fig1}b. 

\begin{figure*}%[t]
\centering
%\hspace*{-2.em}
\includegraphics[width=0.8\textwidth]{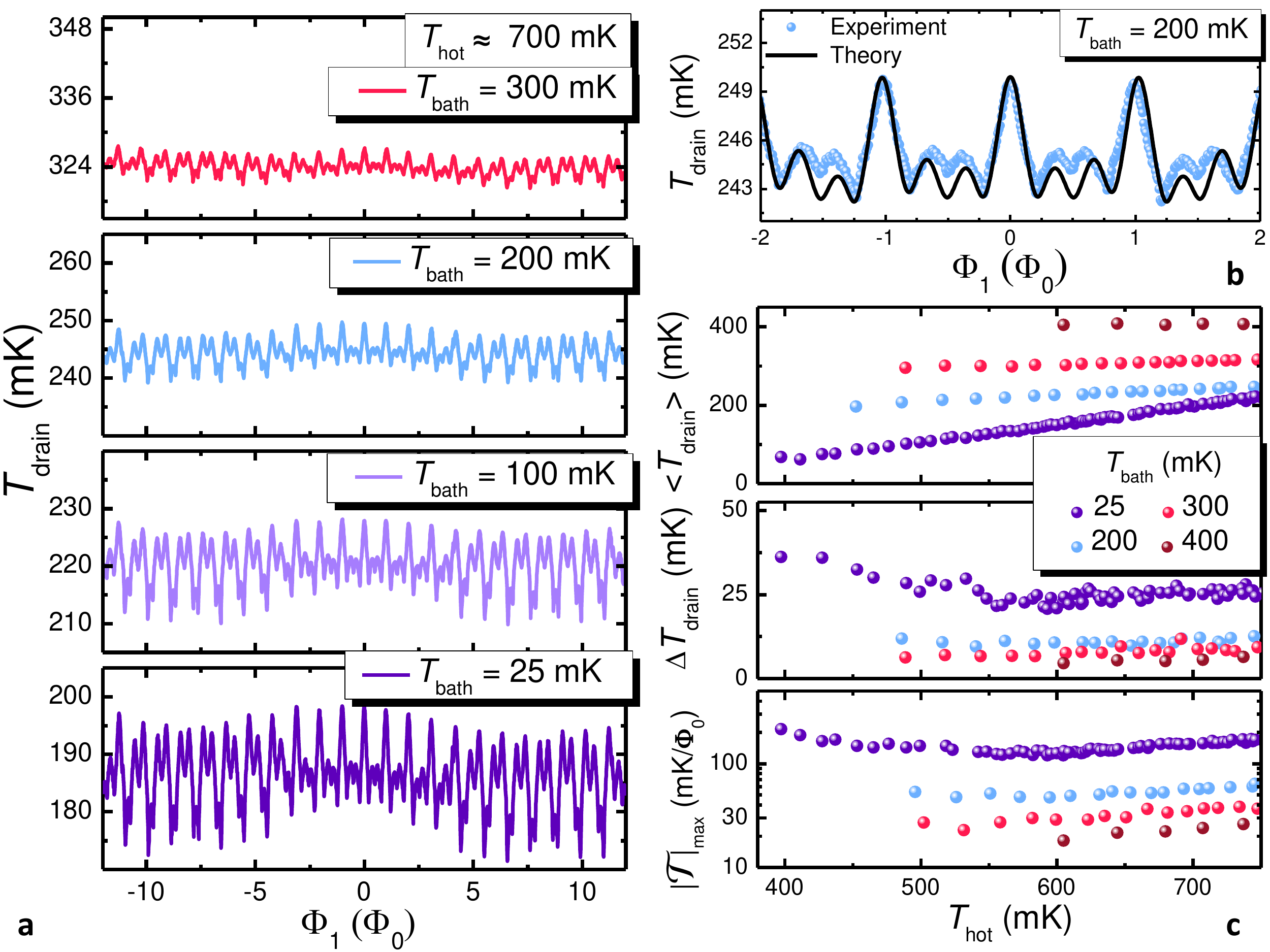}
%\vspace*{-4.ex}
\caption{\textbf{Behaviour of the device at different bath temperatures.} \textbf{a.} Magnetic-flux dependent $T_{\rm drain}$ modulations for different values of $T_{\rm bath}$ at $T_{\rm hot}\simeq 700$ mK. \textbf{b.} Measured $T_{\rm drain}$ (full circles) vs. $\Phi_1$ compared to the theoretical fit (solid black line) for $T_{\rm hot}\simeq 700$ mK and $T_{\rm bath}=200$ mK. \textbf{c} Average drain temperature $\langle T_{\rm drain}\rangle$, maximum temperature modulation amplitude $\Delta T_{\rm drain}=T_{\rm drain}^{\rm max}-T_{\rm drain}^{\rm min}$, and maximum flux-to temperature transfer coefficient $\vert \mathcal{T}\vert_{\rm max}$ vs. $T_{\rm hot}$ for different values of $T_{\rm bath}$. \label{Fig4}} 
%% \vspace*{-3.ex}
\end{figure*} 

Figure~\ref{Fig3}a shows $T_{\rm drain}$ oscillations as a function of the magnetic flux for different values of $T_{\rm hot}$ at $25$ mK. Even at a glance, the thermal magnetic interference pattern perfectly mimics the Josephson critical current one. As $T_{\rm hot}$ becomes higher, the average value of $T_{\rm drain}$ ($\langle T_{\rm drain}\rangle$) raises well above $T_{\rm bath}$ due to the increased heat flow across the structure, while the range of $T_{\rm drain}$ spanned by the oscillations decreases from a maximum of $\sim 40$ mK and tends towards saturation for larger $T_{\rm hot}$.  %As we shall argue, these thermal modulations are indeed generated by three phase-coherent components of the heat current interfering in the core of our device.
We stress that so large temperature modulation amplitudes have never been reported so far, and stem from the design choice and implementation of the Josephson heat modulator.

In order to account for our observations, we have formulated a thermal model outlining all the relevant heat exchange mechanisms present in the structure. The model is sketched in Fig.~\ref{Fig3}b, where the terms $J_{\rm source}$ and $J_{\rm drain}$ denote the quasiparticle heat currents exchanged by S$_1$ with the source and drain electrodes, respectively. The upper branch of the SQUID, as previously mentioned, can release energy to S$_3$ and S$_2$ by means of $J_{\rm probe}$ and $J_{\rm SQUID}$ [see Eq.~\eqref{Jtot}], but only the latter term can be phase-controlled thanks to the double-loop geometry. Finally, electrons in the source and drain electrodes can exchange energy at power $J_{\rm e-ph}$ with lattice phonons residing at $T_{\rm bath}$ (see Methods for details).

The steady-state electronic temperatures $T_1$ and $T_{\rm drain}$ can be calculated for any given $T_{\rm hot}$, $T_{\rm bath}$ and magnetic fluxes $\Phi_1$ and $\Phi_2$ by solving the following system of energy-balance equations:
\begin{align} 
\label{balance1}
%&\dot{Q}_{Heat} = \dot{Q}_{Source}(T_{Hot},T_{Bath})  +\dot{Q}_{e-ph,\, Source}(T_{Hot},T_{Bath}) \\ 
&J_{\rm source} = J_{\rm probe}+J_{\rm SQUID}(\Phi_1,\Phi_2)+J_{\rm drain}, \\ \label{balance2}
&J_{\rm drain} = J_{\rm e-ph,\, drain}.
\end{align}
These equations account for the detailed thermal budget in S$_1$ and drain electrodes by equating the sum of all incoming and outgoing heat currents. In the model we neglect photon-mediated thermal transport owing to poor impedance matching between source and drain electrodes~\cite{Pascal,meschke,schmidt}, as well as electron-phonon coupling in S$_1$ due to its reduced volume and low experimental $T_{\rm bath}$~\cite{Timofeev}.

By imposing conservation of the circulating supercurrent~\cite{MartinezAPLdouble} together with flux quantization [Eqs.~\eqref{Squidsupercurrent}] for both loops, we get
\begin{equation}
J_{\rm SQUID}(\Phi_1,\Phi_2)=(1+r_1+r_2)J_{\rm qp}^{\rm B}-\mathcal{F}_{int}(\Phi_1,\Phi_2)J_{\rm int}^{\rm B},
\label{JSQUID}
\end{equation}
\noindent where $J_{\rm qp}^{\rm B}$ and $J_{\rm int}^{\rm B}$ are detailed in the Methods. From Eq.~\eqref{JSQUID} it is evident that the expected magnetic flux dependence of the heat current is exactly the same obtained for the Josephson critical current. The calculated $T_{\rm drain}$ was fitted to the measured data by using the structure parameters determined from the electrical measurements, and by adding a single phenomenological prefactor to reduce the amplitude of $J_{\rm int}^{\rm B}$ (see Methods for further details).

As shown in Fig.~\ref{Fig3}c, the model provides a remarkable agreement with the experimental data, grasping the relevant heat transport mechanisms in our system. Furthermore, this theoretical picture allows us to estimate the suppression of the phase-coherent component of the heat current in the Josephson thermal modulator, otherwise not attainable from any direct measurement. The fit reveals a maximum contrast $(J_{\rm int}^{\rm max}-J_{\rm int}^{\rm min})/J_{\rm int}^{\rm max}\simeq 99\%$, exactly as in the case of $I_0$. 

Figure~\ref{Fig3}d displays the flux-to-temperature transfer coefficient $\mathcal{T}=\partial T_{\rm drain} /\partial \Phi_1$ for different values of $T_{\text{\rm hot}}$. As previously mentioned, the double-loop geometry allows us to achieve non-trivial heat current-phase relations~\cite{MartinezAPLdouble}, leading to sharp features in $\mathcal{T}$ and to sensitivities which can be as large as $\sim 200$ mK/$\Phi_0$. We note that the latter value is larger by more than a factor of three than that obtained in previous experiments~\cite{GiazottoNature,MartinezNature}. 

The heat modulator dependence on the bath temperature is summarized in Fig.~\ref{Fig4}. Figure~\ref{Fig4}a displays $T_{\rm drain}(\Phi)$ at $T_{\rm hot}\simeq 700$ mK for different values of $T_{\rm bath}$. Even though the distinctive shape remains always well recognizable, the amplitude of the oscillations is progressively depressed by the increasing electron-phonon coupling, which affects mainly the N drain electrode. Nevertheless, our thermal model still provides good theoretical fits for the experimental data, as shown in Fig.~\ref{Fig4}b. Finally, Fig.~\ref{Fig4}c assembles the overall behavior of the heat modulator. Together with $\langle T_{\rm drain}\rangle$, also the maximum temperature modulation amplitude $\Delta T_{\rm drain}=T_{\rm drain}^{\rm max}-T_{\rm drain}^{\rm min}$ and the maximum value of $\vert \mathcal{T}\vert$ are displayed for several values of $T_{\rm hot}$ and $T_{\rm bath}$. As previously noted, for low values of $T_{\rm bath}$ raising $T_{\rm hot}$ produces an initial increase of $\langle T_{\rm drain}\rangle$ joined to a reduction of $\Delta T_{\rm drain}$ and $\vert \mathcal{T}\vert_{\rm max}$. All these quantities tend towards saturation as $T_{\rm hot}$ exceeds $\sim 600$ mK due to the impact of the electron-phonon coupling. This effect is even more noxious as $T_{\rm bath}$ is increased, but not sufficiently strong to completely suppress the magnetic oscillations. The latter are indeed detectable up to $T_{\rm bath}\simeq 400$ mK.

In summary, we have realized a double-loop Josephson thermal modulator that provides a complete phase-engineering of the electronic heat current at the nanoscale. Despite the unavoidable presence of asymmetries in the JJs composing the SQUID, our device offers a suppression of the phase-coherent component of the heat current up to 99\%, leading to a maximum swing in the temperature oscillations of 40 mK and to a flux-to-temperature transfer coefficient as large as 200 mK/$\Phi_0$ at $25$ mK.
Yet, the interferometer demonstrates the perfect correspondence in the manipulation of charge and heat transport, paving the avenue to the ideation of exotic coherent caloritronic nanodevices where thermal currents can be engineered at will. 
In this perspective, our heat modulator could be furnished with two independent superconducting on-chip coils that would provide separate control knobs for the magnetic fluxes.
This would allow to master independently the phase-biasing in the loops~\cite{MartinezAPLdouble}, opening the way to perform closed cycles in the parameter space, and leading to the realization of heat pumps~\cite{Ren} and time-dependent quantum heat engines~\cite{Campisi,Niskanen,Quan}.
This approach can be extremely efficient thanks to the accuracy and high frequency (up to several GHz) at which magnetic fluxes can be driven in the system~\cite{Valenzuela,Niskanen}. On the other hand, our quantum device might represent the building block at the core of thermal splitters~\cite{Bosisio} able to control the amount of energy transferred among several terminals residing at different temperatures. Finally, further studies are needed to extend present results to higher temperatures, exploiting, for instance, high-$T_{\rm c}$ superconductors.

\section*{Methods}
The devices were fabricated with electron-beam lithography and three-angle shadow-mask evaporation of metals onto an oxidized Si wafer through a bilayer resist mask. The evaporations and oxidation were made using an ultra-high vacuum electron-beam evaporator, which allowed us to deposit first 15 nm of Al$_{0.98}$Mn$_{0.02}$ at an angle of 40$^{\circ}$ to form the source and drain N electrodes. Then the sample was exposed to 800 mTorr of O$_{2}$ for 5 minutes to realize the thin insulating layer of AlO$_{\rm x}$ forming the tunnel-barriers in all the SIN junctions present in the system. Afterwards, 20 nm of Al were evaporated at 0$^{\circ}$ to deposit the superconducting island S$_1$, together with the probes used as heaters and thermometers. Another exposition of 200 mTorr of O$_{2}$ for 5 minutes was required to realize the insulating layers of the SIS junctions forming the double loop SQUID and connecting the probe S$_3$ to S$_1$. Finally the sample was tilted at an angle of 30$^{\circ}$ and a deposition of 40 nm of Al was performed to implement the lower branch of the SQUID S$_2$ and the probe S$_3$. The source and drain electrodes have a volume $\mathcal{V}_{\rm source}=1.1\times 10^{-19}$ m$^{-3}$ and $\mathcal{V}_{\rm drain}=7.5\times 10^{-20}$ m$^{-3}$, respectively, whereas the volume of S$_1$ is $\mathcal{V}_{\rm S_1}=2\times 10^{-19}$ m$^{-3}$.

All the measurements have been performed in a filtered dilution refrigerator down to 25 mK. The thermometers were current biased through battery-powered floating sources, while the heaters were piloted with voltage biasing in the range 0-3 mV, corresponding to a maximum power of $\sim 120$ pW injected in the source electrode. Thermometer bias currents were varied from 5 pA to 100 pA in order to maximize the sensitivity in different ranges of temperature and to reduce the effects of self-heating and self-cooling~\cite{GiazottoRev}. Voltage and current were measured with standard room-temperature preamplifiers.
 
In the thermal model, $J_{\rm source}(T_{\rm hot},T_{1})=J_{\rm NIS}(T_{\rm hot},T_{1})$ and $J_{\rm drain}(T_{1},T_{\rm drain})=J_{\rm NIS}(T_{1},T_{\rm drain})$, where $J_{\rm NIS}(T_{\rm N},T_{\rm S})=(2/e^{2}R_{\rm x}) \displaystyle\int_{0}^{\infty}  \! \epsilon \mathcal{N}(\epsilon , T_{\rm S}) \allowbreak [f(\epsilon ,T_{\rm N})-f(\epsilon , T_{\rm S})] \, \mathrm{d}\epsilon$, being $f(\epsilon ,T)=[1+\text{exp}(\epsilon/k_{\rm B} T)]^{-1}$ the Fermi-Dirac distribution, $\mathcal{N}(\epsilon , T)=|\Re[(\epsilon+i\Gamma)/\sqrt{(\epsilon+i\Gamma)^2-\Delta^2(T)}]|$ the smeared (if $\Gamma \neq 0$) normalized Bardeen-Cooper-Schrieffer density of states (BCS DOS) in the superconductor~\cite{Dynes}, $\Delta(T)$ the temperature-dependent energy gap~\cite{Tinkham}, $R_{\rm x}$ the tunnel junction normal-state resistance, $e$ the electron charge and $k_{\rm B}$ the Boltzmann constant. On the other hand, in the expression of $J_{\rm SQUID}$ (see Eq.~\ref{JSQUID}) two components appear: the first accounts for the heat carried by quasiparticles, $J_{\rm qp}(T_{1},T_{\rm bath})=(2/e^{2}R_{\rm x}) \displaystyle\int_{0}^{\infty}  \! \epsilon \mathcal{N}_1(\epsilon , T_{\rm 1})\mathcal{N}_2(\epsilon , T_{\rm bath}) \allowbreak [f(\epsilon ,T_{\rm 1})-f(\epsilon , T_{\rm bath})] \, \mathrm{d}\epsilon$, where $\mathcal{N}_1$ and $\mathcal{N}_2$ are the normalized BCS DOS of the superconductors forming the Josephson junctions. The second component, instead, represents the phase-coherent part of the heat current~\cite{MakiGriffin,Guttman}, which reads $J_{\rm int}(T_{1},T_{\rm bath})=(2/e^{2}R_{\rm x}) \displaystyle\int_{0}^{\infty}  \! \epsilon \mathcal{M}_1(\epsilon , T_{\rm 1})\mathcal{M}_2(\epsilon , T_{\rm bath}) \allowbreak [f(\epsilon ,T_{\rm 1})-f(\epsilon , T_{\rm bath})] \, \mathrm{d}\epsilon$. Here, $\mathcal{M}_{\rm n}(\epsilon,T) =\allowbreak |\Im[-i \Delta_{\rm n}(T)/\sqrt{(\epsilon\allowbreak +i\Gamma_{\rm n})^2-\Delta_{\rm n}^2(T)}]|$ is the Cooper pair BCS DOS in the n-th superconductor, with $\text{n}=1,2$. In the same way, $J_{\rm probe}(T_1,T_{\rm bath})=J_{\rm qp}(T_{1},T_{\rm bath})-J_{\rm int}(T_{1},T_{\rm bath})$. Finally, the electron-phonon coupling in the drain electrode generates the term  $J_{\rm e-ph,drain}(T_{\rm drain},T_{\rm bath})= \allowbreak \Sigma \mathcal{V}_{\rm drain}  \allowbreak (T_{\rm drain}^{6} - \allowbreak  T_{\rm bath}^{6})$, where $\Sigma= \allowbreak 4.5\times \allowbreak 10^{9}$ WK$^{-6}$m$^{-3}$ is the material-dependent electron-phonon coupling constant~\cite{MartinezNature,MartinezNatRect,Maasilta}. 

The theoretical curves in Figs.~\ref{Fig3} and~\ref{Fig4} have been calculated from Eqs.~(\ref{balance1}, \ref{balance2}), using the the measured values of $R_{\rm s}$, $R_{\rm d}$, $R_{\rm p}$, $R_{\rm SQUID}$, $r_1$, $r_2$, $\rho=\Phi_2/\Phi_1$ and $\Gamma\simeq 5\times 10^{-3}\Delta(0)$ as determined from the electrical characterization of the devices.
They were fitted to the experimental data by introducing a single phenomenological factor $A \simeq 0.6 $ reducing the coherent component of the heat current in Eq.~\ref{JSQUID}.
Its origin may be ascribed to fluctuations of the chemical potential in the superconducting island S$_1$, which can induce time-dependent phase evolutions. Indeed, in the thermal measurement setup (see Fig.~\ref{Fig1}b), our device seems more sensitive to such fluctuations because S$_1$ is connected to floating leads S$_2$ and S$_3$ through tunnel barriers. By contrast, in the electrical configuration, we set the phase difference of each JJ present in the device by current-biasing our interferometer. In the latter case, fluctuations turns out to be negligible, as demonstrated by the good agreement between the experimental data and the Ambegoakar-Baratoff prediction (see the inset of Fig.~\ref{Fig2}c).

Furthermore, we estimated numerically the effect of finite self and mutual inductances of the loops.
These inductances renormalize the values of the magnetic fluxes piercing the SQUID~\cite{Tinkham} but their impact on the behaviour of our device turns out to be negligible.

We also checked that the inclusion of a finite Kapitza resistance in the energy-balance equations produces a variation of $\sim 0.5\%$ in the calculated drain temperature.
Similarly, the heat exchange due to the quasiparticle-phonon coupling~\cite{Timofeev} in S$_1$ was also neglected, since this contribution results to be at least ten orders of magnitude smaller than the thermal flux flowing from the source to the drain in the explored range of $T_{\rm hot}$ and $T_{\rm bath}$.  
Finally, the power generated by the electron-photon coupling between the source and the drain~\cite{meschke} has been estimated to be four orders of magnitude smaller than $J_{\rm source}$ and $J_{\rm drain}$, thereby leading to insignificant corrections of our results.

%%%%%%%%%%%%%%%%%%%%%%%%%%%%%%%%%%%%%%%%%%%%%%%%%%%%%%%%%%%%
%%%%%%%%%%%%%%%%%%%%%%   BIBLIOGRAPHY   %%%%%%%%%%%%%%%%%%%%%%%%%%%%%
%%%%%%%%%%%%%%%%%%%%%%%%%%%%%%%%%%%%%%%%%%%%%%%%%%%%%%%%%%%%

\section*{Acknowledgments}
We wish to thank P. Solinas for valuable discussions. The  MIUR-FIRB2013 – Project Coca (Grant No. RBFR1379UX) and the European  Research Council under the European Union’s Seventh Framework Program (FP7/2007-2013)/ERC Grant agreement No. 615187- COMANCHE are acknowledged for partial financial support.

\section*{Author contributions}
C.B. fabricated the samples. A.F., C.B. and S.D.A. performed the measurements. A.F., C.B. and R.B. analysed the data and carried out the simulations. F.G. conceived the experiment. All the authors discussed the results and their implications equally at all stages, and wrote the manuscript.

%%%%%%%%%%%%%%%%%%%%%%%%%%%%%%%%%%%%%%%%%%%%%%%%%%%%%%%%%%%%
%%%%%%%%%%%%%%%%%%%%%%%%%%%%%%%%%%%%%%%%%%%%%%%%%%%%%%%%%%%%
%%%%%%%%%%%%%%%%%%%%%%%%%%%%%%%%%%%%%%%%%%%%%%%%%%%%%%%%%%%%
%%%%%%%%%%%%%%%%%%%%%%%%%%%%%%%%%%%%%%%%%%%%%%%%%%%%%%%%%%%%
%%%%%%%%%%%%%%%%%%%%%%%%%%%%%%%%%%%%%%%%%%%%%%%%%%%%%%%%%%%%
%%%%%%%%%%%%%%%%%%%%%%%%%%%%%%%%%%%%%%%%%%%%%%%%%%%%%%%%%%%%
%FIGURES
%%%%%%%%%%%%%%%%%%%%%%%%%%%%%%%%%%%%%%%%%%%%%%%%%%%%%%%%%%%%

\end{document}